\documentclass{elsart}

\usepackage{amssymb}

\begin{document}

\begin{frontmatter}



\title{Growth of Hydrodynamic Perturbations in Accretion Disks: Possible Route to 
Non-Magnetic Turbulence}


\author{Banibrata Mukhopadhyay, Niayesh Afshordi, Ramesh Narayan}

\address{Institute for Theory and Computation, Harvard-Smithsonian Center for Astrophysics,
60 Garden Street, MS-51, Cambridge, MA 02138, USA}

\begin{abstract}
We study the possible origin of {\it hydrodynamic} turbulence in cold accretion disks such 
as those in star-forming systems and quiescent cataclysmic variables. 
As these systems are expected to have neutral gas, the turbulent viscosity is
likely to be hydrodynamic in origin, not magnetohydrodynamic. Therefore
MRI will be sluggish or even absent in such disks. Although there are no exponentially growing
eigenmodes in a hydrodynamic disk, because of the non-normal nature of the eigenmodes,
a large transient growth in the energy is still possible, which may enable the system to switch to a
turbulent state. For a Keplerian disk, we estimate that the energy will grow by a factor of $1000$
for a Reynolds number close to a million.

\end{abstract}

\end{frontmatter}


\section{INTRODUCTION}

In the context of Keplerian disks, while hydrodynamic simulations seem 
to confirm the absence of turbulence
in the non-linear limit \cite{hbsw}, the
Magneto-Rotational Instability \cite{chandra}, discovered by Balbus and
Hawley \cite{bh} within ionized accretion flows, confirms the existence of
MHD turbulence. This helps to understand the origin of shear stress or ``viscosity".
However the MRI driven instability fails to operate in
disks with small ionization fraction. Examples of such
systems are proto-planetary disks, outskirts of AGN accretion
disks \cite{gam} etc.. As a result, the route to turbulence and subsequently
accretion in neutral disks has remained one of the outstanding
puzzles in modern astrophysics.

Laboratory experiments of Taylor-Couette systems
(which are similar to accretion systems) seem to indicate
that, the flow is unstable to turbulence for Reynolds numbers larger than a
few thousand \cite{richard}, even for subcritical
systems. Based on this, Longaretti \cite{long} concludes that 
the absence of turbulence in previous simulations \cite{hbsw} is due
to their small effective Reynolds number.

The fundamental reason for subcritical turbulence may be the result of
{\it definite frequency} modes not being orthogonal in a shear flow. Therefore, a 
suitably tuned linear combination of modes can
show an arbitrarily large transient energy growth\footnote{Growth is defined as the
ratio of energy at a particular
instant to that at the beginning.}. This transient growth may possibly lead to
sustained {\it hydrodynamic turbulence} for large enough Reynolds numbers.

Here we analyze this transient growth phenomenon in the framework of accretion disks.
We present the set of scaling relations for the transient
growth as a function of the Reynolds number.

\section{BASIC EQUATIONS AND SCALING RELATIONS}

Let us define the shear frequency $2A$ and the vorticity frequency
$2B$ as, $2A = -q\Omega, \, 2B = (2-q)\Omega$.
Then the momentum balance equations give
\begin{eqnarray}
\hskip-1.2cm
{du\over dt}= 2\Omega v - {\partial \hat{p}\over \partial x}
+\nu {\nabla}^2 u,\,
{dv\over dt} = -2B u - {\partial \hat{p}\over \partial y}
+\nu {\nabla}^2 v,\,
{dw\over dt}= - {\partial \hat{p}\over \partial z}
+\nu {\nabla}^2 w,
\label{zmmtm}
\end{eqnarray}
along with the incompressibility relation ${\partial u/\partial x} + {\partial v/\partial y} + {\partial
w/\partial z} = 0$, where $\hat{p}=P/\rho$.  Symbols $u,v,w$ and $
\Omega$ are three velocity components and angular frequency
respectively, $\nu, P, \rho$ are dynamical viscosity, pressure and density of the system respectively
and $q$ is a parameter which controls the flow pattern (whether a Keplerian disk, a constant angular
momentum disk or plane Couette flow). 
The Lagrangian time derivative $d/dt$ is, 
${d/dt} = \partial/{\partial t} -q\Omega x\partial/{\partial y}$.
Here $x$ varies
from $-L$ to $+L$, and $y$ and $z$ from $0$ to $2\pi/k_y$ and $2\pi/k_z$ respectively, where
$k_x$, $k_y$, $k_z$, are the corresponding components of wave-vectors.
The boundary conditions are $u=v=w=0$ at $x=\pm L$.

\subsection{Constant Angular Momentum Disks and Plane Couette Flow}

It is clear from (\ref{zmmtm}) that the set of equations is very symmetrical for a constant
angular momentum ($q=2$) and plane Couette ($\Omega=0$) flow, except that $x$ and $y$ are interchanged.
Therefore the growth is identical in the two cases. As
the previous analytical/numerical studies (e.g. \cite{bf,man}) showed that the growth is 
maximum for vertical perturbations ($k_y=0$), here we consider such perturbation.
Noting that the Reynolds number is, $R = |2A|L^2/\nu$,
we find that the maximum growth in energy for a given $k_z$ is
\begin{eqnarray}
G_{max}(k_z,R) = \left(\frac{u^2(t)+v^2(t)+w^2(t)}{u^2(0)+v^2(0)+w^2(0)}\right)_{\rm optimum}
={R^2k_z^2L^2\,e^{-2}\over \left[{1\over2}\pi^2
+k_z^2L^2\right]^3}.
\end{eqnarray}
Maximizing this over $k_z$, we find that the optimum wavevector is
$k_zL = \pi/2 = 1.57$.
The maximum growth factor and the corresponding time are
\begin{eqnarray}
\nonumber
G_{max}(R) = 0.82\times10^{-3}R^2,\,\,
|2A|t_{max}(R) = 0.13 R.\\
\end{eqnarray}
Clearly the growth scales as $R^2$ and therefore can be very large even for modest $R$.
For example, when $R=1200$, the above relations give 
$G_{max}(R)>1000$, which is probably large enough to induce non-linear feedback and turbulence.

\subsection{Keplerian Disk}

Here we study eqns. (\ref{zmmtm}) for $q=1.5$. As the earlier analysis \cite{man} hinted that $k_z=0$
generates the best growth, we consider such perturbations.

We consider a plane wave that is frozen into the fluid and is sheared
along with the background flow.  If the flow
starts at time $t=0$ with initial wave-vector $\{k_{xi}, k_y\}$, 
the $k_x$ at later time is given by 
$k_x(t) = k_{xi} + q\Omega k_yt$.
The maximum energy growth is then approximately 
\begin{eqnarray}
G_{max}(k_{xi},k_y,R) \sim {k_{xi}^2L^2 \over (1.7)^2+k_y^2L^2}
\exp\left(-{2\over 3R}{k_{xi}^3L^2\over k_y}\right).
\end{eqnarray}
when $k_{x,min}L=1.7$.
Maximizing this with respect to $k_{xi}$ and $k_y$, we obtain
\begin{eqnarray}
G_{max}(R) \sim 0.13R^{2/3}, \,\,\, |2A|t_{max}(R) \sim 0.88 R^{1/3}.
\end{eqnarray}
Also maximizing with respect to $k_{xi}$, keeping $k_y$ and $R$ fixed, we obtain
\begin{eqnarray}
\hskip-1.0cm
G_{max}(k_y,R)= {(k_yL)^{2/3} \over (1.7)^2+(k_yL)^2}
e^{-2/3}R^{2/3},\,\,|2A|t_{max}(k_y,R)= (k_yL)^{-2/3} R^{1/3}.
\end{eqnarray}
We see that the maximum growth scales as $R^{2/3}$. Though this rate of increase is less than 
that for a $q=2$ disk and plane Couette flow, still, at a large enough $R$, the growth can
become large, and may cause turbulence in a Keplerian disk. For example, 
for $R=10^6$, $G_{max}(R)> 1000$, which by analogy with a $q=2$ disk may be enough to cause
turbulence.

\section{CONCLUSION}

We have shown that significant transient growth of
perturbations is possible in a Keplerian flow. Although the system does not 
have any unstable eigenmodes, because of the non-normal nature of the eigenmodes a
significant level of transient energy growth is possible at a large Reynolds number for
appropriate choice of initial conditions. 
We argue that a plausible critical Reynolds number for sustaining hydrodynamic turbulence
in a Keplerian disk is
$\sim 10^4-10^6$ (for detailed discussions, see \cite{man}).
We base this on the expectation that once the growth crosses the threshold, secondary instabilities of
various kinds, such as the elliptical instability (see e.g. \cite{ker}), might
serve as a possible route to self-sustained turbulence.
However, it remains to be verified that
these instabilities are present. Also, even if they are present, one will need to investigate whether
they lead to non-linear feedback and 3-dimensional turbulence.
 
This work was supported in part by NASA grant NAG5-10780 and NSF grant
AST 0307433.



\end{document}